\documentclass[11pt]{article}

\usepackage{enumerate}
\usepackage{graphicx}
\usepackage{amsmath}
\newcommand{\Feynp}[1]{#1\kern-0.42em/}

\def\ep{\varepsilon}

\def\half{\frac{1}{2}}
\def\quater{\frac{1}{4}}

\def\O{\mathcal{O}}

\def\lr#1{\left(#1\right)}
\def\slr#1{\left[#1\right]}

\def\trl#1{\textrm{Tr}\lr{#1}}
\def\trsl#1{\textrm{Tr}\slr{#1}}
\def\avg#1{\left\langle #1\right\rangle}

\begin{document}
\fontfamily{cmr}\fontsize{11pt}{14pt}\selectfont
\def \CMP {{Commun. Math. Phys.}}
\def \PRL {{Phys. Rev. Lett.}}
\def \PL {{Phys. Lett.}}
\def \NPBProc {{Nucl. Phys. B (Proc. Suppl.)}}
\def \NP {{Nucl. Phys.}}
\def \RMP {{Rev. Mod. Phys.}}
\def \JGP {{J. Geom. Phys.}}
\def \CQG {{Class. Quant. Grav.}}
\def \MPL {{Mod. Phys. Lett.}}
\def \IJMP {{ Int. J. Mod. Phys.}}
\def \JHEP {{JHEP}}
\def \PR {{Phys. Rev.}}
\def \JMP {{J. Math. Phys.}}
\def \GRG{{Gen. Rel. Grav.}}
\begin{titlepage}
\null\vspace{-62pt} \pagestyle{empty}
\begin{center}
\rightline{CCNY-HEP-13/01}
\rightline{January 2013}
\vspace{1truein} {\Large\bfseries
Random matrix approach to scalar fields on fuzzy spaces}\\
\vskip .1in
{\Large\bfseries
~}\\
\vspace{6pt}
\vskip .1in
{\Large \bfseries  ~}\\
\vskip .1in
{\Large\bfseries ~}\\
{\large\sc Juraj Tekel}\\
\vskip .2in
{\itshape Physics Department\\
City College of the CUNY\\
New York, NY 10031}\\

\vskip .1in
{\itshape Physics Department\\
Graduate Center of the CUNY\\
New York, NY 10016}\\

\vskip .1in
\begin{tabular}{r l}
E-mail:
&{\fontfamily{cmtt}\fontsize{11pt}{15pt}\selectfont jtekel@gc.cuny.edu}\\

\end{tabular}

\fontfamily{cmr}\fontsize{11pt}{15pt}\selectfont
\vspace{.8in}
\centerline{\large\bf Abstract}
\end{center}

We formulate theory of interacting scalar field on the fuzzy sphere as a random matrix model. We then analyze the expectation values of observables of the theory in the large $N$ limit and we demonstrate that the eigenvalue distribution of the matrix $M$ remains the polynomially deformed Wigner semicircle. We also compute distributions involving the matrix Laplacian of $M$ and we show that the correlation between the eigenvalues of these two is different from the free field case.

\end{titlepage}
\pagestyle{plain} \setcounter{page}{2}

\section{Introduction}

\!Matrix models have established a firm place in the modern physics. Starting with pioneering work of Wigner in description of the spectrum of heavy nuclei, they have emerged in areas of string theory as finite approximations to Riemann surfaces \cite{string}, condensed matter system as the Calogero model \cite{guhr} or chaos in quantum systems \cite{chaos}. Fuzzy spaces are non-commutative spaces with a finite dimensional underlying Hilbert space \cite{ncgeo}. It is possible to describe them by finite dimensional matrices and they become their commutative counterpart in the limit of these matrices being very large. The physical motivation of using fuzzy spaces is in regulating the divergences without breaking the isometries of the space-time, which is especially appealing to quantum gravity considerations \cite{douglas}. Fuzzy spaces also arise as brane solutions in string theory and in $M$-theory \cite{taylor}.

In \cite{our}, it was shown that matrix models and fuzzy spaces are related. One can generalize the standard Gaussian matrix ensembles by addition of matrix Laplacian. This procedure is motivated by the the scalar field theories on fuzzy spaces and introduces new observables involving matrix derivatives. Averages of some of these were considered. This was then used to show, that the eigenvalue distributions of the matrix and its Laplacian are correlated and the joint distribution was computed. From physical point of view, such ensembles provide new computational tools for the fuzzy theories, for example to compute distributions of various observables.

In this paper, we present some further work along these lines. In the first part, we compute expectation values and derive the distributions of more general observables in such defined random matrix ensembles. This shows that this framework is very robust and can be used beyond simple observables of the form $M^m B^b$, where $B$ is the matrix Laplacian of $M$. We also show, that there is a different scaling of the terms in the action, under which the contribution of the mass term survives the limit of very large matrices. Moreover, the correlation between the distributions of $M$ and $B$ then remains finite for any form of the kinetic term in this limit. In the second part, we concentrate on the matrix ensemble motivated by an interacting field theory. We show that, as was the case before, the distribution of eigenvalues of $M$ is altered only by a rescaling of the variable. We then investigate the properties of observables involving $B$. We show that the distribution of the eigenvalues of $B$ and the joint distribution of $M$ and $B$ are connected and we compute each fo these up to the second order in the $MB$ correlation. We also show that the interaction brings some new features into the problem, which were not present in the free case.

Some aspects of this problem were approached by other authors from different points of view. Authors of \cite{ocon} treated the Laplacian term as a perturbation and integrated out the angular degrees of freedom. The eigenvalue problem was then solved using the standard methods. After the corresponding approximation is made, our results presented here agree with the results computed using this method in \cite{samann}. In \cite{stein}, the kinetic term and added interaction term are treated exactly, but the eigenvalue distribution is said to have the same destiny as in the free case and different aspects of the results are considered. Here, we present more complete calculation of the distribution and we also give calculation of expectation values of observables involving the Laplacian.

\section{Results for the free theory}\label{sec2}

\!We will consider the Euclidean theory of a real scalar field on the fuzzy sphere governed by the general kinetic term action
\begin{eqnarray}\label{freeS}
	S_0(M)=\half \trl{M\mathcal K M}+\half \mu^2 \trl{M^2}.
\end{eqnarray}
We will introduce interaction terms later and for now we will consider only the free theory. The standard Laplacian kinetic term is given by $\mathcal K M=[L_\alpha,[L_\alpha,M]]$, where $L_\alpha$ are the generators of the $N$ dimensional representation of $SU(2)$. We denote $\mathcal K M=B$. We further introduce a basis for fields on the fuzzy sphere in terms of $N\times N$ matrices
\begin{eqnarray}
	T^l_m \ \ \ , \ \ \ l=0,1,\ldots,N-1 \ \ \ ,\ \ \ m=-l,-l+1,\ldots,l-1,l \ \ \ ,
\end{eqnarray}
normalized as $\trl{T^l_m\,T^{l'}_{m'}}=\delta^{ll'}\delta_{mm'}$. We can expand the matrix $M$ in terms of this basis as
\begin{eqnarray}
	M=\sum_{l,m}c^l_m T^l_m.
\end{eqnarray}
The non-interacting action is diagonal in this basis and the correlator of two components of $M$ is $\avg{c^l_m\,c^{l'}_{m'}}=\delta^{ll'}\delta_{mm'}G(l)$ where the propagator $G(l)$ depends on the form of the kinetic term. For example for the standard kinetic term it is $G(l)=1/(\mu^2+l(l+1))$.

We define the following two-point functions
\begin{eqnarray}
	\avg{(MM)_{ij}}=f\delta_{ij} \ \ \ , \ \ \ \avg{(BB)_{ij}}=g\delta_{ij} \ \ \ , \ \ \ \avg{(MB)_{ij}}=h\delta_{ij}.
\end{eqnarray}
Also $f=\avg{\trl{MM}}/N$ and similarly for $g$ and $h$. In fact, these three correlators are all the information we need about the theory and we do not need to know the form of $\mathcal K$ explicitly. This also means that with a proprer choice of $\mathcal K$, we could work with the theory on different fuzzy spaces.

\subsection{Previous results}

\!Here, we very briefly summarize the results of \cite{our}, which we are going to use later. In this section, we will normalize the distributions of $M$ and $B$ to have unit radius. This eliminates some cumbersome factors of $2$ in final formulas. In the sections to follow, we will however change this normalization to eliminate factors of two from equations we will work with and to follow the standard convention. This should be kept in mind when comparing results from these sections \ref{sec2} and \ref{sec4}.

We define the normalized correlators
\begin{eqnarray}
	W_{m,b} = \frac{1}{N}\avg{  \trsl{ \lr{\frac{M}{2\sqrt{f}}}^m \lr{\frac{B}{2\sqrt{g}}}^b }}.
\end{eqnarray}
We can write down, in the large $N$ limit, the following recursion rules for them
\begin{eqnarray}\label{recWmb}
4W_{m,b} &=& \sum_{p =0}^{m-2} W_{p,0} W_{m-2-p, b} 
+ \gamma \, \sum_{p =0}^{b-1} W_{m-1, b -1-p} W_{0,p} \ \ \ \ \ m\geq1,\nonumber\\
4W_{m,b} &=& \sum_{p =0}^{b-2} W_{0,p} W_{m, b-p-2} 
+ \gamma \, \sum_{p =0}^{m-1} W_{m -1- p, b -1} W_{p,0} \ \ \ \ \ b\geq1,
\end{eqnarray}
where $\gamma=h/\sqrt{fg}$. These are results of explicit Wick contractions of matrices and the planarity of corresponding diagrams. Defining the generating function
\begin{eqnarray}
	\phi (t,s ) = \sum_{m, b =0}^\infty \, W_{m,b} \, t^m\, s^b
\end{eqnarray}
these recursion rules become equations for $\phi(t,s)$, which can be solved as
\begin{eqnarray}
	\phi(t,0)\equiv\phi(t)&=&2\frac{1-\sqrt{1-t^2}}{t^2}=\frac{2}{1+\sqrt{1-t^2}} \ \ \ , \ \ \ \phi(0,s)=\phi(s) \ \ \ ,\\
	\phi(t,s)&=&\frac{\phi(s)(\phi(t)}{1 - \quater\gamma ts \phi(s)(\phi(t)}.\label{achach}
\end{eqnarray}
Inverting this generating function we arrive at the final formula for the distribution function
\begin{eqnarray}\label{jointdistr}
	\rho(x,y)=\rho(x)\rho(y)\frac{1-\gamma^2}{(1-\gamma^2)^2-4\gamma(1+\gamma^2)xy+4\gamma^2(x^2+y^2)},
\end{eqnarray}
where $\rho(x)=2\sqrt{1-x^2}/\pi$ is the Wigner semicircle distribution generated by $\phi(t)$. This distribution is positive for $\gamma^2<1$ and in the limit $\gamma\to\pm1$ becomes $\rho(x)\delta(x-y)$, so the two matrices are completely correlated/anti-correlated, as expected.

Therefore the distribution of eigenvalues of the unnormalized $M$ has radius $2\sqrt f$, distribution of eigenvalues of $B$ has radius $2 \sqrt g$ and the correlation between the two is given by the above formula.

\subsection{Joint distribution of eigenvalues of three matrices}

\!Since the two point functions of the matrices $M$ and $B$ were the only relevant quantities for computation of the joint distribution of the previous section, if we define 'linear' matrices $A$ as
\begin{eqnarray}\label{one}
	A_1,A_2=\sum_{m,l} c_m^l f_{1,2}(l,m)T_m^l,
\end{eqnarray}
the whole procedure will go through basically intact and we recover the same results. Namely the joint distribution of the eigenvalues of $A_1$ and $A_2$ is going to be given by
\begin{eqnarray}
	\rho(x,y)=\rho(x)\rho(y)\frac{1-\gamma_{12}^2}{(1-\gamma_{12}^2)^2-4\gamma_{12}(1+\gamma_{12}^2)x y-4\gamma_{12}^2(x^2+y^2)}
\end{eqnarray}
and $\gamma_{12}=\avg{Tr(A_1 A_2)}/\sqrt{a_1 a_2}$, where $a_i=\avg{TrA_i^2}$. The distributions of eigenvalues of matrices $A_1,A_2$ have radius $2 \sqrt{a_{1,2}}$. In terms of the original definition we have
\begin{eqnarray}
	a_{1,2}=\frac{1}{N}\sum_{l=0}^{N-1} G(l)\sum_{m=-l}^{l}f_{1,2}^2(l,m).
\end{eqnarray}
We now turn to computation of the joint distribution of three matrices of the form (\ref{one}), i.e. we look for $\rho(x,y,z)$ such that
\begin{eqnarray}
	W_{a,b,c}=\frac{1}{N}\avg{\trsl{\lr{\frac{A_1}{2\sqrt{a_1}}}^a \lr{\frac{A_2}{2\sqrt{a_2}}}^b \lr{\frac{A_1}{2\sqrt{a_1}}}^c}}=\int dx dy dz\ x^a y^b z^c \rho(x,y,z).
\end{eqnarray}
Our approach will be again to write down recursion rules for the moments $W_{a,b,c}$ coming from the Wick contractions. There are three different types of contractions
\begin{eqnarray}
	\avg{\trl{A_1^a A_2^b A_3^c}}&\to&
	 \avg{\trl{A_1^{p-2}}}a_1 N\avg{\trl{A_2^{b}A_3^c}}\nonumber\\
	&\to&\avg{\trl{A_1^{a-1}A_2^p}}\sqrt{a_1a_2}\gamma_{12} N\avg{\trl{A_2^{b-p-1}A_3^c}}\nonumber\\
	&\to&\avg{\trl{A_1^{a-1}A_2^b A_3^p}}\sqrt{a_1a_3}\gamma_{13} N\avg{\trl{A_3^p}}.
\end{eqnarray}
Each contraction splits the diagram into two parts and planarity condition forbids contracting matrices in different parts. Summing over all possible contractions we obtain
\begin{eqnarray}
4	W_{a,b,c}&=&\sum_{p=0}^{a-2}W_{p,0,0} W_{a-p-2,b,c} + \gamma_{12} \sum_{p=0}^{b-1}W_{a-1,p,0} W_{0,b-p-1,c}+\nonumber\\&+&\gamma_{13}\sum_{p=0}^{c-1}W_{0,0,p} W_{a-1,b,c-p-1} \ \ \ \ \ a\geq1.
\end{eqnarray}
This rule holds only for $a\geq1$, because for $a=0$ there is no matrix $A_1$ to do the contraction with. The quantities $W_{0,b,c}$ are moments of two point distribution of matrices $A_2$ and $A_3$ and are given by (\ref{recWmb}). In this recursion rule they enter as the initial condition.

Considering contractions of the other groups of matrices we obtain two more equations
\begin{eqnarray}
4	W_{a,b,c}&=&\sum_{p=0}^{b-2}W_{0,p,0} W_{a,b-p-2,c} + \gamma_{12} \sum_{p=0}^{a-1}W_{p,0,0} W_{a-p-1,b-1,c}+\nonumber\\&+&\gamma_{23}\sum_{p=0}^{c-1}W_{a,0,p} W_{0,b-1,c-p-1} \ \ \ \ \ b\geq1,\\
4	W_{a,b,c}&=&\sum_{p=0}^{b-2}W_{0,0,p} W_{a,b,c-p-2} + \gamma_{13} \sum_{p=0}^{a-1}W_{0,p,0} W_{a-p-1,b,c-1}+\nonumber\\&+&\gamma_{23}\sum_{p=0}^{b-1}W_{a,p,0} W_{0,b-1-p,c-1} \ \ \ \ \ c\geq1.
\end{eqnarray}
We solve these by defining the generating function
\begin{eqnarray}
	\phi(t,s,u)=\sum_{a,b,c}W_{a,b,c}t^a s^b u^c
\end{eqnarray}
and by rewriting the recursion relations as equations for $\phi(t,s,u)$
\begin{eqnarray}
	4\Big(\phi(t,s,u)-\phi(0,s,u)\Big)&=&t^2\phi(t,0,0)\phi(t,s,u)+\gamma_{12}ts\phi(t,s,0)\phi(0,s,u)+\nonumber\\&+&\gamma_{13}tu\phi(0,0,u)\phi(t,s,u),\label{abc1}\\
	4\Big(\phi(t,s,u)-\phi(t,0,u)\Big)&=&s^2\phi(0,s,0)\phi(t,s,u)+\gamma_{12}ts\phi(t,0,0)\phi(t,s,u)+\nonumber\\&+&\gamma_{23}su\phi(t,0,u)\phi(0,t,u),\label{abc2}\\
	4\Big(\phi(t,s,u)-\phi(0,s,u)\Big)&=&u^2\phi(0,0,u)\phi(t,s,u)+\gamma_{13}tu\phi(0,s,0)\phi(t,s,u)+\nonumber\\&+&\gamma_{23}su\phi(t,s,0)\phi(0,t,u).
\end{eqnarray}
We require $\phi(0,0,0)=W_{0,0,0}=1$ as a normalization condition. Now, choosing appropriate variables to be zero, we can solve these equations. For example in the first relation, setting $s=u=0$ we get
\begin{eqnarray}
	4\phi(t,0,0)-4=t^2 \phi^2(t,0,0),
\end{eqnarray}
which is the same equation we have arrived at in the two matrix case and has solution
\begin{eqnarray}
	\phi(t,0,0)\equiv\phi(t)=\frac{2}{1+\sqrt{1-t^2}}.
\end{eqnarray}
Similarly for the case of $\phi(0,s,0)$ and $\phi(0,0,u)$.

Setting $u=0$ in (\ref{abc1}) we get equation for $\phi(t,s,0)$ which solves again for the formula obtained in the previous section
\begin{eqnarray}
	\phi(t,s,0)\equiv\phi(t,s)=\frac{4\phi(s)}{4-t^2\phi(t)-\gamma t s \phi(s)}=\frac{4\phi(s)}{\frac{4}{\phi(t)}-\gamma t s \phi(s)}=\frac{\phi(s)\phi(t)}{1-\quater\gamma_{12} t s \phi(t)\phi(s)}.
\end{eqnarray}
We could obtain the same quantity from (\ref{abc2}) by setting $u=0$ and this obviously yields the same result. Next, we set $s=0$ in the first equation and obtain
\begin{eqnarray}
	\phi(t,0,u)=\frac{\phi(t)\phi(u)}{1- \quater\gamma_{13} t u \phi(t)\phi(u)}
\end{eqnarray}
and the same way for the rest of the functions.

Plugging these into one of the original equations we get the final formula for the generating function
\begin{eqnarray}\label{gen3a}
	\phi(s,t,u)=\frac{\phi(s)\phi(t)\phi(u)}{\Big(1-\quater s t\gamma_{12}\phi(s)\phi(t)\Big)\Big(1-\quater t u\gamma_{13}\phi(t)\phi(u)\Big)
	\Big(1-\quater s u\gamma_{23}\phi(s)\phi(u)\Big)}.
\end{eqnarray}
With no surprise, this is the formula we obtain from either of the three equations.

We will now describe a general method that can be used to invert generating functions of this form to obtain the corresponding distribution. We present the proof in the appendix \ref{proof}. If the generating function is expressed as
\begin{eqnarray}
	\phi(t_1,\ldots,t_n)=f\big(t_1\phi(t_1),\ldots,t_n\phi(t_n)\big)\phi(t_1)\ldots\phi(t_n),
\end{eqnarray}
then the corresponding distribution is given by
\begin{eqnarray}\label{panati}
	\rho(x_1,\ldots,x_n)=\sum_{\ep_i=\pm1}\ep_1\ldots\ep_n F(e^{i\ep_1\theta_1},\ldots,e^{i\ep_n\theta_n}),
\end{eqnarray}
where
\begin{eqnarray}\label{panati2}
	F(z_1,\ldots,z_n)=\lr{\prod_{j=1}^n\frac{z_j}{i\pi}}f(2z_1,\ldots,2z_n)
\end{eqnarray}
and $x_n=\cos\theta_n$. Using this for the generating function (\ref{gen3a}) gives a very complicated formula of the form
\begin{eqnarray}\label{rho3}
	\rho(x,y,z)=\rho(x)\rho(y)\rho(z)\times \tilde\rho_3(x,y,z).
\end{eqnarray}
Explicit formula for $\tilde\rho_3(x,y,z)$ is given in the appendix \ref{explicit}.

However couple of important observations can be made. In the case $\gamma_{12,23,13}=0$ the factor becomes $1$, in the case of two of the three $\gamma$'s vanishing the factor becomes the appropriate function to give $\rho(x,y,z)=\rho(x,y)\times \rho(z)$, if $\gamma_{12,23,31}\to1$ we get $\rho(x,y,z)\to \rho(x)\delta(x-y)\delta(x-z)$ and if $\gamma_{12,23}\to-1,\gamma_{23}\to1$ we get$\rho(x,y,z)\to \rho(x)\delta(x+y)\delta(x+z)$, i.e. fully correlated or anti-correlated distributions as expected and finally in the case of $\gamma_{12}=\gamma_{13}=\gamma,\gamma_{23}\to1$ we obtain $\rho(x,y,z)\to \rho(x,y)\delta(y-z)$.

\subsection{$MBMB$ joint distribution}

\!To get the four point joint distribution of matrices $MBMB$, we need to compute the following quantities
\begin{eqnarray}\label{gen4mb}
	W_{a,b,c,d}=\frac{1}{N}\avg{\trsl{\lr{\frac{M}{2\sqrt f}}^a \lr{\frac{B}{2\sqrt g}}^b \lr{\frac{M}{2\sqrt f}}^c \lr{\frac{B}{2\sqrt g}}^d}}.
\end{eqnarray}
Using the same explicit Wick contractions and planarity of the diagrams we find the following large $N$ recursion rule
\begin{eqnarray}
	4W_{a,b,c,d}&=&\sum_{p=0}^{a-2}W_{p,0,0,0} W_{a-p-2,b,c,d} + \gamma \sum_{p=0}^{b-1}W_{0,p,0,0} W_{a-1,b-p-1,c,d}+\nonumber\\
	&+&\sum_{p=0}^{c-1}W_{0,b,p,0} W_{a-1,0,c-p-1,d} + \gamma \sum_{p=0}^{b-1}W_{a-1,0,0,p} W_{0,b,c,d-p-1}
\end{eqnarray}
for $a\geq1$. We have considered all the possible contractions of the first matrix of the $M^a$ part. Introducing the generating function
\begin{eqnarray}
	\phi(t,s,u,v)=\sum_{a,b,c,d} W_{a,b,c,d}\ t^a s^b u^c v^d
\end{eqnarray}
this becomes
\begin{eqnarray}
	4\Big(\phi(t,s,u,v)-\phi(0,s,u,v)\Big)&=&t^2 \phi(t,0,0,0) \phi(t,s,u,v)+\gamma t s \phi(0,s,0,0) \phi(t,s,u,v)+\nonumber\\
	&+&t u \phi(0,s,u,0) \phi(t,0,u,v) + \gamma t v \phi(t,0,0,v)\phi(0,s,u,v).
\end{eqnarray}
Now if we consider contractions of matrices from other parts of the diagram, we get three more equations for the generating function, namely
\begin{eqnarray}
	4\Big(\phi(t,s,u,v)-\phi(t,0,u,v)\Big)&=&s^2\phi(0,s,0,0) \phi(t,s,u,v)+ \gamma s u\phi(0,0,u,0) \phi(t,s,u,v)+\nonumber\\
	&+&s v\phi(0,0,u,v) \phi(t,s,0,v) + \gamma s t \phi(t,s,0,0)\phi(t,0,u,v),\\
	4\Big(\phi(t,s,u,v)-\phi(t,s,0,v)\Big)&=&u^2\phi(0,0,u,0) \phi(t,s,u,v)+\gamma u v \phi(0,0,0,v) \phi(t,s,u,v)+\nonumber\\
	&+&u t \phi(t,0,0,v) \phi(t,s,u,0) + \gamma u s \phi(0,s,u,0)\phi(t,s,0,v),\\
	4\Big(\phi(t,s,u,v)-\phi(t,s,u,0)\Big)&=&v^2 \phi(0,0,0,v) \phi(t,s,u,v)+\gamma v t \phi(t,0,0,0) \phi(t,s,u,v)+\nonumber\\
	&+&v s\phi(t,s,0,0) \phi(0,s,u,v) +\gamma  v u \phi(0,0,u,v)\phi(t,s,u,0).
\end{eqnarray}
We again have $\phi(0,0,0,0)=W_{0,0,0,0}=1$. Then, choosing appropriate variables to be zero, we can solve these for the generating function pretty much the same way we did in the case of three matrices, with the same results for $\phi(t,0,0,0),\phi(t,s,0,0),\phi(t,s,u,0)$ and the rest of the combinations. We can make the notation more compact by defining
\begin{eqnarray}
	\psi(t,s)=1-\quater\gamma_{ts} t s \phi(t)\phi(s),
\end{eqnarray}
where $\gamma_{ts}$ is the correlation parameter between the matrices corresponding to the variables $t$ and $s$. This way
\begin{eqnarray}
	\phi(\tau,\sigma)&=&\frac{\phi(\tau)\phi(\sigma)}{\psi(\tau,\sigma)},\\
	\phi(\tau,\sigma,\rho)&=&\frac{\phi(\tau)\phi(\sigma)\phi(\rho)}{\psi(\tau,\sigma)\psi(\tau,\rho)\psi(\sigma,\rho)},
\end{eqnarray}
where $\tau,\sigma,\rho$ are any of $t,s,u,v$. Finally (\ref{gen4mb}) becomes
\begin{eqnarray}
	\phi(t,s,u,v)&=&\frac{\phi(t)\phi(s)\phi(u)\phi(v)}{\psi(t,s)\psi(t,v)\psi(u,v)\psi(s,u)\psi(s,v)\psi(t,u)}\nonumber\\&&\ \ \ \times\ \Big[1-\frac{tsuv}{16}\phi(t)\phi(s)\phi(u)\phi(v)\Big].
\end{eqnarray}
Using the method described in detail in the previous section, this yields the four point distribution of the form
\begin{eqnarray}\label{rho4}
	\rho(x,y,z,w)=\rho(x)\rho(y)\rho(z)\rho(w)\times\tilde\rho_4(x,y,z,w),
\end{eqnarray}
with explicit formula for $\tilde\rho_4(x,y,z,w)$ is given in the appendix \ref{explicit}. Here we just observe that the factor is $1$ when $\gamma=0$, in the case of $\gamma\to1$ we get
 $\rho(x,y,z,w)\to\rho(x)\delta(x-y)\delta(z-w)\delta(x-w)$ and in the case of $\gamma\to-1$ $\rho(x,y,z,w)\to\rho(x)\delta(x+y)\delta(z+w)\delta(x+w)$, i.e. the variables correctly correlate or anti-correlate.

\section{Mass rescaling and the correlation $\gamma$}

\!The large $N$ structure of correlation parameter $\gamma$ was discussed in \cite{our}. It was shown that if $G(l)$ goes like $l^\alpha$ for large $l$, correlation $\gamma$ can vanish or tend to a constant value  in the large $N$ limit, depending on the value $\alpha$. Especially for the Laplacian kinetic term, i.e. $\alpha=-2$ the correlation vanishes as $1/\log N$. In this section, we show that there is a different scaling of the kinetic and the mass terms of the action and that upon a rescaling of $\mu$ correlation $\gamma$ is always finite.

The $MM$ correlator is given by
\begin{eqnarray}
	f=\frac{1}{N}\avg{\trl{MM}}=\frac{1}{N}\sum_{l=0}^{N-1}(2l+1)G(l).
\end{eqnarray}
In the case of Laplacian kinetic term, the large $N$ limit of this expression is
\begin{eqnarray}\label{expf}
	f=\int_0^{1-1/N} dx \frac{2Nx+1}{Nx(Nx+1)+\mu^2},
\end{eqnarray}
which has the advertised $\log N /N$ behavior which leads to logarithmic vanishing of $\gamma$. If we however rescale the mass $\mu^2\to N^2 \tilde \mu^2$, we will find that $f$ now depends polynomially on $N$. Namely
\begin{eqnarray}
	f=\int_0^1 dx \frac{2Nx+1}{Nx(Nx+1)+N^2 \tilde \mu^2}=\frac{1}{N}\log\lr{1+\frac{1}{\tilde\mu^2}}.
\end{eqnarray}
Similar calculation then yields also
\begin{eqnarray}
	h&=&N\slr{1-\tilde\mu^2\log\lr{1+\frac{1}{\tilde\mu^2}}}\\,
	g&=&N^3\slr{\half - \tilde\mu^2+\tilde\mu^4\log\lr{1+\frac{1}{\tilde\mu^2}}}
\end{eqnarray}
and finally finite
\begin{eqnarray}
	\gamma=\frac{\slr{	1-\tilde\mu^2\log\lr{1+\frac{1}{\tilde\mu^2}}	}}	{\sqrt{\log\lr{1+\frac{1}{\tilde\mu^2}}\slr{\half - \tilde\mu^2+\tilde\mu^4\log\lr{1+\frac{1}{\tilde\mu^2}}}}}.
\end{eqnarray}
Note that this is finite in the limit of very large $\tilde \mu$ and tends to $\sqrt3/2$, which is the correlation in the case of no kinetic term.

The same line of attack works also in the case of a general kinetic term $\mathcal K$. We have already assumed, that $\mathcal K T^l_m$ depends only on $l$ and lets assume that for large $l$, this is proportional to $l^\alpha$. We therefore need to rescale $\mu\to N^\alpha \tilde \mu^2$, so that for large $l$, the mass term in the propagator does not get suppressed. The same procedure as before then yields
\begin{eqnarray}
	f&=&N^{1-\alpha}\frac{1}{\tilde\mu^2}\left._2F_1\right.\lr{\frac{2}{\alpha},1;1+\frac{2}{\alpha};-\frac{1}{\tilde\mu^2}},\\
	h&=&N \slr{1-\left._2F_1\right.\lr{\frac{2}{\alpha},1;1+\frac{2}{\alpha};-\frac{1}{\tilde\mu^2}}},\\
	g&=&N^{1+\alpha}\slr{\frac{2}{2+\alpha}-\tilde \mu^2+\tilde\mu^2\left._2F_1\right.\lr{\frac{2}{\alpha},1;1+\frac{2}{\alpha};-\frac{1}{\tilde\mu^2}}},
\end{eqnarray}
where $\left._2F_1\right.$ is the ordinary hypergeometric function. These clearly give a finite $\gamma$.

Therefore an appropriate rescaling of the mass makes the mass term always unsuppressed and keeps the correlation parameter $\gamma$ finite for any kinetic term. This is important when one sets to apply these results in the field theory, since then we want all the terms to survive the large $N$ limit.

\section{Interacting theory}\label{sec4}

\!We are now ready to introduce the interaction the free action (\ref{freeS}). We will consider a quartic interaction potential
\begin{eqnarray}\label{Sint}
	S_{int}=\tilde g \trl{M^4} \ \ \ , \ \ \ \tilde g=g/N.
\end{eqnarray}
As mentioned in the introduction, the case without the kinetic term is well known \cite{brezin},\cite{leshouches} and the result is a polynomial correction to the Wigner semicircle distribution.

Expanding the interaction part in power series in $\tilde g$ yields for an average of an observable $\O(M)$
\begin{eqnarray}\label{intO}
	\avg{\O}=\frac{1}{Z}\sum_{a=0}^\infty \frac{(-\tilde g)^a}{a!}\int dM e^{-S_0(M)} \O(M)\trl{M^4}^a,
\end{eqnarray}
where
\begin{eqnarray}
	Z=\sum_{n=0}^\infty \frac{(-\tilde g)^n}{n!}\int dM e^{-S_0(M)}\trl{M^4}^n.
\end{eqnarray}
So we see that evaluating the average in the interacting theory can be done using averages of the free theory. We just need to pick correct diagrams that contribute to the expectation value on the RHS of (\ref{intO}). The diagrams containing vacuum bubbles, i.e. parts, where some of the vertexes from $\trl{M^4}^a$ contract only among themselves, will be canceled by the $1/Z$ factor. therefore we can write
\begin{eqnarray}\label{intO2}
	\avg{\O}=\sum_{a=0}^\infty \frac{(-\tilde g)^a}{a!}\avg{\O(M)\trl{M^4}^a}_{0,con},
\end{eqnarray}
where the subscript will indicate that we consider only diagrams that do not contain disconnected vacuum bubbles and that the contractions are to be taken using the free theory measure.

Now we also see the motivation for the $1/N$ factor in the definition of the coupling constant $\tilde g$. Since each trace in $\avg{f(M)\trl{M^4}^a}$ raises large $N$ dependence of this expression by one, expectation values in the previous sum are going to be all of the same order and all the terms will contribute in the large $N$ limit.

Before we proceed with computation of the eigenvalue distribution of $M$, let us stress one point. The contractions in (\ref{intO2}) are done using the free measure. But we have already seen that in the free theory, the kinetic term only rescaled the radius of the original distribution. And therefore we expect the same in the interacting case, namely that all the distributions of the $M^4$-theory with no kinetic term will survive also in the full theory, only with a rescaled variable.

\subsection{Eigenvalue distribution of the matrix $M$}

\!To compute the distribution of eigenvalues of $M$ we need to compute the moment generating function $\phi_1(t)$. The potential is even and therefore odd moments vanish and we need to compute $\avg{Tr\ M^{2m}}$. From (\ref{intO2}) we see that we need to investigate quantities of the form
\begin{eqnarray}\label{F}
	F^{2m}_a=\frac{1}{N^{1+a}}\avg{
	\trsl{\lr{
	\frac{M}{\sqrt f}}^{2m}
	}
	\trsl{
	\lr{\frac{M}{\sqrt f}}^4}\ \overset{a\ \textrm{times}}{\ldots\ldots\ldots}\ \trsl{\lr{\frac{M}{\sqrt f}}^4}
	}_{0,con}.
\end{eqnarray}
Expressions
\begin{eqnarray}\label{Fam}
	F_{2m}=\sum_{a=0}^\infty\frac{(-g)^a}{a!}F^{2m}_a
\end{eqnarray}
are then going to be finite and will give $2m$ point correlators of $M$'s in the interacting theory. Let us stress again that in this section, recursion rules for the free correlators change a little, due to a different normalization of the distribution of $M$.
\begin{figure}\label{fig1}                    
\begin{center}
\includegraphics[height=60mm]{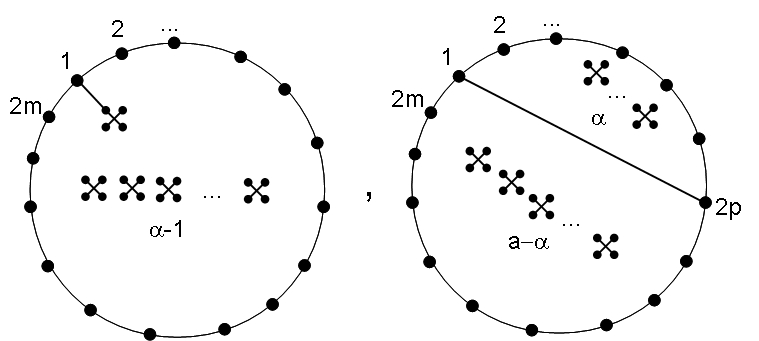}   
\end{center}
\vspace{-2mm}
\caption{Two types of terms contributing to (\ref{recintMraw}).}
\end{figure}
We will now write down the recursion rules for $F^{2m}_a$ in a very similar fashion we did in previous cases. The contributing planar diagrams are going to be of the form of $2m$ points on a circle with four-point vertexes inside this circle. The number of vertexes is $a$ and the point and vertexes connect with no lines intersecting. If we look at one point on the circle, there are two different types of contractions we can make. With a vertex, three legs of the vertex become effectively three new points on the circle, which now has $2m+2$ points, but there are only $a-1$ vertexes left and there are $4a$ different legs that we can connect the first point to. Or with a different point, the circle splits into two circles, each new circle can have different number of vertexes in it, one circle $\alpha$ and the other $a-\alpha$, with a proper combinatorial factor. This procedure is illustrated in the figure \ref{fig1}. We need to be a little careful. If there are no points on any circle to connect with, we can not have any vertexes in this circle, as this would produce a non-connected diagram. therefore we need to set $F^0_a=0$ for any $a\neq0$ and $F^0_0=1$.

Summing over all possible contractions, we obtain
\begin{eqnarray}\label{recintMraw}
	\avg{\trl{M^{2m}}\slr{\trl{M^4}}^a}=4a f N\avg{\trl{M^{2m+2}}\slr{\trl{M^4}}^{a-1}}+\nonumber\\+ f N\sum_{p=1}^{m}\sum_{\alpha=0}^a \binom{a}{\alpha} \avg{\trl{M^{2(p-1)}}\slr{\trl{M^4}}^\alpha} \avg{\trl{M^{2(m-p)}}\slr{\trl{M^4}}^{a-\alpha}}
\end{eqnarray}
and using the definition (\ref{F}) this becomes
\begin{eqnarray}\label{recFma}
	F^{2m}_a=4a F^{2m+2}_{a-1}+\sum_{p=1}^{m}\sum_{\alpha=0}^a \frac{a!}{\alpha!(a-\alpha)!} F^{2(p-1)}_\alpha F^{2(m-p)}_{a-\alpha}.
\end{eqnarray}
Immediately we note, that if we define $\tilde F^{2m}_a=F^{2m}_a/a!$ this simplifies into
\begin{eqnarray}
	\tilde F^{2m}_a=4\tilde F^{2(m+1)}_{a-1}+\sum_{p=0}^{m-1}\sum_{\alpha=0}^a \tilde F^{2p}_\alpha \tilde F^{2(m-1-p)}_{a-\alpha}.
\end{eqnarray}
Multiply the recurrence relation by $(-g)^a$ and sum over $a=1$ to $\infty$ to obtain
\begin{eqnarray}
	F_{2m}-F^{2m}_0=-4gF_{2(m+1)}+\sum_{p=0}^{m-1}\slr{F_{2p}F_{2(m-1-p)}-F^{2p}_0F^{2(m-1-p)}_0}.
\end{eqnarray}
We know from the free theory that $F^{2m}_0=c_n$ and using the identity for the Catalan numbers $c_n=\sum_{p=0}^{n-1}c_p c_{n-1-p}$ the extra terms cancel. Note that this is consequence of the free case recursion rules, and similar terms will cancel for this reason also later. We therefore get
\begin{eqnarray}\label{recF}
		F_{2(m+1)}=\frac{1}{4g}\slr{\sum_{p=0}^{m-1}F_{2p}F_{2(m-1-p)}-F_{2m}}.
\end{eqnarray}
This expression holds however only for $m\geq1$, so we need to specify $F_0$ and $F_2$. From the definition of $F^0_a$ it is clear that $F_0=1$, consistent with the normalization of the distribution. $F_2$ is essentially the dressed propagator of the interacting theory and for a while lets go further without specifying it.

We now define the moment generating function $\phi_1(t)=\sum t^m F_m=\sum t^{2m}F_{2m}$, multiply the previous formula by $t^{2(m+1)}$ and sum over $m=1$ to $\infty$ to obtain
\begin{eqnarray}
	\phi_1(t)-1-t^2F_2=\frac{1}{4g}\slr{t^4\phi_1^2(t)-t^2(\phi_1(t)-1)}
\end{eqnarray}
or
\begin{eqnarray}
	t^4\phi_1^2-\lr{4g+t^2}\phi_1+\lr{4g+t^2+4gt^2F_2}=0
\end{eqnarray}
which gives
\begin{eqnarray}\label{genphi1}
	\phi_1(t)=\frac{4g+t^2-\sqrt{(4g+t^2)^2-4t^4(4g+t^2)-16 F_2 gt^6}}{2t^4}.
\end{eqnarray}
So we are left to specify the two point function $F_2$. 

At this point, we are going to take $F_2$ to be the expression obtained by the standard methods \cite{brezin},\cite{leshouches}. This formula is explicitly
\begin{eqnarray}\label{bs}
	F_2=\frac{(1+48g)^{3/2}-1-72g}{864g}=\frac{2}{3}a^2(4-a^2),
\end{eqnarray}
where $a^2=(\sqrt{1+48g}-1)/24g$. This might seem that we are assuming something we are trying to prove, but this is not the case. We assume that the initial condition we are about to use in (\ref{recF}) is the way we expect it to be, i.e. the same as for the case of no kinetic term. We then compute the generating function and if it turns out to be the same, we conclude that this assumption leads to all the correlators being of the form as in the theory with no kinetic term.

Before we show that this is indeed the case, let us go back to (\ref{recFma}) for a while. We can use this recursion rule to generate a lot of terms $\tilde F^{2m}_a$ and after some trial and error using the integer factorization of the terms we can guess the formula for in the following form
\begin{eqnarray}\label{guess}
	\tilde F^{2m}_a=12^a\lr{\frac{(2m)!}{m!(m-1)!}}\lr{\frac{(m+2a-1)!}{(m+a+1)!a!}}.
\end{eqnarray}
And not surprisingly this is indeed the formula of $g$ expansion of the $2m$-point correlator given in the appendix of \cite{brezin}. It would be interesting to see, whether one can extract $F_2$ in a closed form directly from the recursion rule (\ref{recFma}) without solving for $\tilde F^{2m}_a$. Or one could try to prove that (\ref{guess}) solves the recursion rule (\ref{recFma}) by explicit computation or some inductive method. We have attempted this, but the problem is more complicated and we will proceed with assumption (\ref{bs}).

We plug this formula for $F_2$ into the the generating function (\ref{genphi1}), which after some algebra can be brought into the form
\begin{eqnarray}
	\phi_1(t)=\frac{t^2+4g}{2t^4}-\frac{1}{t^2}\lr{\half + 4 g a^2+\frac{2g}{t^2}}\sqrt{1-4 a^2 t^2}.
\end{eqnarray}
We have recovered the standard generating function for the distribution of $M$. Now, the discussion goes along the usual lines, the resolvent is
\begin{eqnarray}\label{resol}
	\omega(\lambda)=\frac{1}{\lambda} \phi_1(1/\lambda)=\int_{-2a}^{2a}dx\frac{\rho_1(x)}{\lambda-x},
\end{eqnarray}
which yields for the distribution
\begin{eqnarray}
	\frac{1}{\pi}\lr{\half+4 g a^2+2 g x^2}\sqrt{4a^2-x^2}.
\end{eqnarray}
This is the polynomial deformation to the Wigner semicircle distribution, with the radius $2a$. Looking back at (\ref{F}), the variable in the case of the unscaled matrix is $x/\sqrt f $ and we need to replace $g\to f^2 g$,\footnote{To see this better, we should carry the explicit factors of $f$ in the calculation. Such calculation would yield an extra factor of $f^{m+2a}$ in (\ref{guess}). From the definition of $\phi_1(t)$ we can see that $f^{m}$ rescales $t$ by $\sqrt{f}$ and from (\ref{Fam}) we can see that $f^{2a}$ rescales $g$ by $f^2$.} i.e. the final formula is
\begin{eqnarray}\label{velkyvysledok}
	\rho_1(x)=\frac{1}{\pi}\lr{\frac{1}{2f}+4 g f a^2+2 g x^2}\sqrt{4a^2f-x^2}.
\end{eqnarray}

Therefore starting from the recurrence relation (\ref{recFma}) we have been able to recover the result of the distribution of eigenvalues of the random matrix ensemble with weight (\ref{Sint}). In the next section, we will generalize this approach to different observables of the interacting theory. After using the explicit formula (\ref{expf}) for $f$ we see, that the expression (\ref{velkyvysledok}) reduces to the previous result in \cite{samann}, where a polynomial deformation of the Wigner distribution was obtained also.\footnote{To do this, one has to introduce a parameter $\ep$ in front of the kinetic term, do the expansion of $f$ in powers of $\ep$ and take $\ep=1$ at the end of the calculation.}

\subsection{Eigenvalue distribution of the matrix $B$ and the joint $MB$ distribution}

\!As in the free case, the theory now includes new observables involving the matrix $B=\mathcal K M$. In the free case, this matrix followed the same distribution as the underlying matrix $M$. Now, the situation is going to be different, since the interaction involves only the matrix $M$.

We will discuss the distribution of eigenvalues of $B$ and the joint distribution for $M$ and $B$. These two are going to be connected, since contractions of $B$ with matrix $M$ in the interaction vertex are going to turn even pure $B$ correlators into mixes $MB$ ones. Define
\begin{eqnarray}
	G^{2b}_a&=&\frac{1}{N^{1+a}}\avg{\trsl{\lr{\frac{B}{\sqrt g}}^{2b}}\trsl{
	\lr{\frac{M}{\sqrt f}}^4}\ \overset{a\ \textrm{times}}{\ldots\ldots\ldots}\ \trsl{\lr{\frac{M}{\sqrt f}}^4}}_{0,con},\\ W^{m,b}_a&=&\frac{1}{N^{1+a}}\avg{\trsl{\lr{\frac{M}{\sqrt f}}^{m}\lr{\frac{B}{\sqrt g}}^b}\trsl{
	\lr{\frac{M}{\sqrt f}}^4}\ \overset{a\ \textrm{times}}{\ldots\ldots\ldots}\ \trsl{\lr{\frac{M}{\sqrt f}}^4}}_{0,con}
\end{eqnarray}
and
\begin{eqnarray}
	G_{2b}=\sum_{a=0}^\infty\frac{(-g)^a}{a!}G^{2b}_a=\sum_{a=0}^\infty(-g)^a\tilde G^{2b}_a\ \ \ , \ \ \ W_{m,b}=\sum_{a=0}^\infty\frac{(- g)^a}{a!}W^{m,b}_a=\sum_{a=0}^\infty(-g)^a\tilde W^{m,b}_a.
\end{eqnarray}
Using the same approach as before, it is now quite easy to write down the recurrence rule for $\tilde G^{2m}_a$
\begin{eqnarray}\label{recG}
	\tilde G^{2b}_a=4\gamma \tilde W^{3,2b-1}_{a-1}+\sum_{p=0}^{b-1}\sum_{\alpha=0}^a \tilde G^{2p}_\alpha \tilde G^{2(b-1-p)}_{a-\alpha},
\end{eqnarray}
where the first term comes from the contraction of $B$ with a vertex. And again, this holds only for $b\geq1$. It is also not too difficult to write down the recursion rules for $\tilde W$'s. Here, we can obtain two different recursions, considering contraction of the first $M$ matrix or the first $B$ matrix. These two are
\begin{eqnarray}\label{recWmba}
	\tilde W^{m,b}_a=\sum_{p=0}^{m-2}\sum_{\alpha=0}^a \tilde F^p_\alpha \tilde W^{m-p-2,b}_{a-\alpha}
	+\gamma \sum_{p=0}^{b-1}\sum_{\alpha=0}^a \tilde  G^p_\alpha \tilde W^{m-1,b-1-p}_{a-\alpha}
	+4\tilde W^{m+2,b}_{a-1}
\end{eqnarray}
holding for $m\geq1,a\geq1$ and
\begin{eqnarray}
	\tilde W^{m,b}_a=\sum_{p=0}^{b-2}\sum_{\alpha=0}^a \tilde G^p_\alpha \tilde W^{m,b-p-2}_{a-\alpha}
	+\gamma \sum_{p=0}^{m-1}\sum_{\alpha=0}^a \tilde F^p_\alpha \tilde W^{m-1-p,b-1}_{a-\alpha}
	+4\gamma \tilde W^{m+3,b-1}_{a-1}
\end{eqnarray}
for $b\geq1,a\geq1$. Note that for $b=0$ in the first and $m=0$ in the second we recover the recursion rules for $\tilde F$'s and $\tilde G$'s respectively. Also in these expression $\tilde W^{m,b}_0$ are considered as initial values, given by the recursion rules of the free case from the previous sections.

Setting $m=1$ in the first of the recursion we get
\begin{eqnarray}
	4\gamma\tilde W^{3,b}_{a-1}&=&\gamma \tilde W^{1,2b-1}_a-\gamma^2 \sum_{k=0}^{b-1}\sum_{\alpha=0}^a \tilde  G^{2k}_\alpha \tilde G^{2(b-1-k)}_{a-\alpha},
\end{eqnarray}
where we have used the fact, that $W^{0,b}_a\equiv G^b_a$ is nonzero only for even $b$. Using this in the the recursion rule for $\tilde G$ (\ref{recG}) we find
\begin{eqnarray}
	\tilde G^{2b}_a=\gamma \tilde W^{1,2b-1}_{a}+(1-\gamma^2)\sum_{p=0}^{b-1}\sum_{\alpha=0}^a \tilde G^{2p}_\alpha \tilde G^{2(b-1-p)}_{a-\alpha}
\end{eqnarray}
and
\begin{eqnarray}
	G_{2b}=\gamma W_{1,2b-1}+(1-\gamma^2)\sum_{p=0}^{b-1} G_{2p} G_{2(b-1-p)}.
\end{eqnarray}
We again define the generating function $\phi_2(s)=\sum s^{2n} G_{2n}$, which yields
\begin{eqnarray}
	\phi_2(s)-1=(1-\gamma^2)s^2\phi^2_2(s)+\gamma s \slr{\sum_{b=1}^\infty W_{1,2b-1}s^{2b-1}}=(1-\gamma^2)s^2\phi^2_2(s)+\gamma s \underbrace{\slr{\sum_{b=0}^\infty W_{1,b}s^{b}}}_{W_1(s)},
\end{eqnarray}
where we have used the fact that $W_{1,b}$ vanishes for even $b$. We see that this is very different from the equation for $\phi_1(t)$, as expected due to the different role of $M$ and $B$ in the interaction. The solution is given by
\begin{eqnarray}\label{rho2}
	\phi_2(s)=\frac{1-\sqrt{1-4s^2(1-\gamma^2)(1+\gamma s W_1)}}{2(1-\gamma^2)s^2}.
\end{eqnarray}
From the condition $\phi_2\to\phi_0$ in the limit of $g\to0$ we recover that in this limit $W_1\to\gamma s \phi_0^2$. We will prove that this is indeed the case shortly.

In the following, we will drop the argument of $\phi_1$ and $\phi_2$. It will be understood that the former is always functions of $t$ and the latter function of $s$.

From the recursion rules for $\tilde W$'s, we can derive an equation for the generating function of the two point distribution
\begin{eqnarray}
	\phi(t,s)=\sum_{m,b}t^m s^b W_{m,b}.
\end{eqnarray}
To do this, we express (\ref{recWmba}) as
\begin{eqnarray}
	W_{m,b}-W^{m,b}_0&=&\sum_{p=0}^{m-2}\slr{F_p W_{m-p-2,b}-F^p_0W^{m-p-2,b}_0}+\gamma\sum_{p=0}^{b-1}\slr{G_pW_{m-1,b-1-p}-G^p_0W^{m-1,b-p-1}_0}\nonumber\\&-&4gW_{m+2,b}.
\end{eqnarray}
The terms with a subscript $0$ will cancel, since they follow the modified relations for the free quantities (\ref{recWmb}). Continuing the procedure we arrive at
\begin{eqnarray}\label{eqn2}
	\phi(t,s)-\phi_2=t^2\phi_1\phi(t,s)+\gamma ts \phi_2\phi(t,s)-\frac{4g}{t^2}\slr{\phi(t,s)-\phi_2-t W_1(s)-t^2 W_2(s)},
\end{eqnarray}
where we have denoted $W_2(s)=\sum_{b=0}^\infty W_{2,b}s^b$. A similar equation can be derived from the second recurrence rule
\begin{eqnarray}\label{eqn1}
	\phi(t,s)-\phi_1=s^2\phi_2\phi(t,s)+\gamma ts \phi_1\phi(t,s)-\frac{4g\gamma s}{t^3}\slr{\phi(t,s)-\phi_2-tW_1(s)-t^2 W_2(s)}.
\end{eqnarray}
Before, we have used multiple equations for the generating function as a consistency check. However by now we trust our procedure enough to using these two together to reduce the number of unknown functions to one. Doing this, we get the final formula for $\phi(t,s)$
\begin{eqnarray}\label{totalbrutal}
	\phi(t,s)=\frac{\phi_2\Big(t \phi_1-\gamma s \phi_2\Big)}{t- \gamma s \phi_2+\gamma s t W_1}.
\end{eqnarray}
So to compute $\phi_2(s)$ and $\phi(t,s)$ we need to specify the function $W_1(s)$.

Let us stress here, that the coefficient $\gamma$ in these expressions is the very same correlation as in expressions of section \ref{sec2}. This shows, that even though we consider different matrix ensemble, it still knows about the underlying fuzzy sphere it wast built on, which is therefore encoded solely in the kinetic term of (\ref{freeS}).

\subsection{Leading order $W_1(s)$ in $\gamma$}

\!In the case of $\tilde F^{2m}_a$, we have been able to guess the solution of the recursion rule. However for the case of $\tilde W^{1,2b-1}_a$ the situation is more involved and no easy guess is possible. Moreover, simple analysis of the integer factorization of the first couple of terms shows, that such guess might be very difficult and the explicit formula much more complicated than a simple product of factorials.

Therefore we will compute $\tilde W^{1,2b-1}_a$ only in the leading order in $\gamma$ to get the first nontrivial contribution due to the $MB$ contraction. In such case, the contributing diagrams have only one $MB$ contraction, leading to one factor of $\gamma$. Let us stress that the results we will obtain are still exact in $g$ and we make no assumption about the magnitude of the coupling.

\begin{figure}\label{fig2}                    
\begin{center}
\includegraphics[height=60mm]{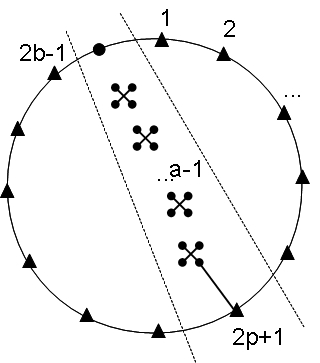}   
\end{center}
\vspace{-2mm}
\caption{Triangles represent matrices $B$. With only one $MB$ contraction, the diagram splits into two parts of matrices $B$ and contractions among $M$'s.}
\end{figure}
As figure \ref{fig2} illustrates, we get the following expression for the diagrams with one $MB$ contraction
\begin{eqnarray}
	\tilde W^{1,2b+1}_a=4\gamma \tilde W^{4,0}_{a-1}\sum_{p=0}^b c_p c_{b-p}=\gamma \tilde F^2_a c_{b+1},
\end{eqnarray}
where we have used the recursion rule (\ref{recFma}). This holds for $a\geq$. For $a=0$ we simply have $W^{1,2b+1}_0=\gamma\sum c_p c_{b-p}=\gamma c_{b+1}=\gamma F^2_0 c_{b+1}$. This then yields
\begin{eqnarray}
	W_{1,2b+1}=\sum_{a=0}^\infty (-g)^a W^{1,2b+1}_a=\gamma c_{b+1} F_2
\end{eqnarray}
and
\begin{eqnarray}\label{W1gamma1}
	W_1=\sum_{b=0}^\infty s^b W_{1,b}=\sum_{b=0}^\infty s^{2b+1} W_{1,2b+1}=\gamma F_2\frac{\phi_0(s)-1}{s}=\gamma s \phi_0^2(s) F_2.
\end{eqnarray}
We use this result in the equation (\ref{rho2}) to obtain the leading contribution to the distribution $\phi_2$
\begin{eqnarray}
	\phi_2(s)=\phi_0(s)+\gamma^2 \phi_0(s)\frac{1-\phi_0(s)+s^2 \phi_0^2(s) F_2}{2-\phi_0(s)}=\phi_0(s)+\gamma^2(F_2-1)\phi_0(s)\frac{s^2\phi_0^2(s)}{1-s^2\phi_0^2(s)}.
\end{eqnarray}
In the limit of $g\to0$, $F_2$ becomes $1$, the whole extra contribution vanishes and $\phi_2(s)=\phi_0(s)$.

Expanding (\ref{totalbrutal}) up to second order in $\gamma$ yields 
\begin{eqnarray}
	\phi(t,s)&=&\phi_0(s)\phi_1(t)+\gamma s \phi_0^2(s) \frac{\phi_1(t)-1}{t}+\nonumber\\&+&\gamma^2\slr{\phi_1(t)(F_2-1)\phi_0(s)\frac{s^2\phi_0^2(s)}{1-s^2\phi_0^2(s)}+
	\frac{(1-F_2 t^2)\phi_1(t)-1}{t^2}s^2\phi_0^3(s)}.
\end{eqnarray}

We now need to invert these to obtain the distributions. We can use the standard approach of defining the resolvent (\ref{resol}) and then using the discontinuity equation\footnote{For more details, see e.g. \cite{leshouches}}
\begin{eqnarray}
	\rho(x)=-\frac{1}{2\pi i}\slr{\omega(x+i\ep)-\omega(x-i\ep)}.
\end{eqnarray}
But it is easier to use a fact mentioned in the appendix \ref{proof} that the non-singular part of $\phi(t)/t^n$ generates the distribution $x^n\rho(x)$, where $\rho(x)$ is generated by $\phi(t)$. This way we find out that
\begin{eqnarray}
	s^2\phi_0^3(s)=\frac{(1-s^2)\phi_0(s)-1}{s^2},
\end{eqnarray}
generates $y^2(1-y^2)\rho_0(y)$ and similarly for $((1-F_2 t^2)\phi_1(t)-1)/t^2$. After some algebra, we find final result for the distribution of the eigenvalues of matrix $B$ and the joint distribution of $M$ and $B$, valid up to second order in the correlation $\gamma$ to be
\begin{eqnarray}
	\rho_2(y)&=&\rho_0(y)\lr{1+\gamma^2(1-F_2)\frac{y^2-2}{y^2-4}},\nonumber\\
	\rho(x,y)&=&\rho_1(x)\rho_2(y)\Big(1+\gamma x y + \gamma^2 x^2(1-F_2 x^2)y^2(1-y^2)\Big).
\end{eqnarray}
As in the case of the distribution of $M$, we have to change $g\to g f^2,x\to x/\sqrt f, y\to y \sqrt g$ to get the distributions of the unscaled matrices. Note that both of these become the appropriate free expressions but there is something new in the second formula. One could guess that the interacting result for $\rho(x,y)$ would be just the free case with the one matrix marginals replaced by the interacting expressions. The extra factor of $F_2$ shows that this is not the case.

\section{Conclusions and Outlook}

The main results we present are twofold. First, we have shown that it is possible to compute distributions of more general observables in the matrix ensemble discussed in \cite{our}. Also, we show how rescaling of the mass puts all the terms on the same footing in the large $N$ limit. Second, we have shown that the matrix ensemble corresponding to the interacting scalar field on the fuzzy sphere does reproduce the distribution of eigenvalues of the model without the kinetic term, which is the polynomially deformed Wigner semicircle. The radius of the distribution gets renormalized by a factor of $\sqrt f$ as was the case for the free field ensemble and this therefore seems to be a generic feature. We have obtained the equations for the distribution of matrix $B$ and the joint distribution of $M$ and $B$ and we have solved these up to an unknown function $W_1$. We have then computed the first non-trivial contribution, which enabled us to compute these distributions to the second order in $\gamma$. The results we have obtained show, that the interaction introduces a novel features into the joint distribution of $M$ and $B$ beyond changing the one matrix marginals.

There are two points that deserve further treatment. First of all, the proof that the formula (\ref{guess}) does indeed solve the recursion rule (\ref{recFma}), or equivalently recovering the expression (\ref{bs}) for $F_2$ directly from the recursion. Also more systematic treatment of the $\gamma$ expansion of $\phi_2(s)$ and $\phi(t,s)$ is needed and could lead to better understanding and possibly complete solution for $W_1(s)$.

With the results presented and possibly some future progress there are several issues to explore. The presented results treat the kinetic term non-perturbatively, therefore it is going to be interesting to analyze the phase diagram of the theory and to compare the findings with previous numerical work \cite{numerical}. It has been also shown that a modification of the kinetic term can remove the tadpole diagrams responsible for the UV/IR mixing \cite{uvir}. Our method is well suited for such modification and one will be able to study signatures of this removal in phase diagram of the theory.

\vskip .2in
{\bf Acknowledgments.} 
I am grateful to V.P. Nair, A.P. Polychronakos and O. Bud\'{a}\v{c} for a lot of fruitful discussions.
This work was supported by U.S.\ National Science
Foundation grant PHY-0855515
and by a PSC-CUNY grant.

\appendix

\section{Proof of the formula (\ref{panati})}\label{proof}

We will prove the one dimensional case. The generalization to more dimensions is then straightforward. The presented proof is a modification of approach used in \cite{our}.

Let us have a generating function of the form
\begin{eqnarray}
	\Phi(t)=f\big(t\phi(t))\phi(t),
\end{eqnarray}
where $\phi(t)=2/(1+\sqrt{1-t^2})$ is the generating function for the Wigner semicircle distribution. We expand this as
\begin{eqnarray}
	\Phi(t)=\sum_{n=0}^\infty a_n t^n \phi^{n+1}(t).
\end{eqnarray}
Using the explicit formula for $\phi(t)$ we obtain
\begin{eqnarray}\label{prrrd}
	t^n \phi^{n+1}(t)=\sum_{k=0}^{[n/2]}\binom{n+1}{2k+1}\lr{\frac{2}{t}}^n(1-t^2)^k\phi(t)-2\sum_{k=0}^{[(n-1)/2]}\binom{n}{2k+1}\lr{\frac{2}{t}}^n(1-t^2)^k.
\end{eqnarray}
Now follows a crucial observation. If a generating function $g(t)$ corresponds to the distribution $\rho(x)$, then the non-singular part of $g(t)/t^n$ generates $x^n\rho(x)$ for any $n\geq0$. In previous expression, the second term contains only singular terms. However the left hand side is clearly non-singular and therefore the second term cancels the singular part of the first term but does not contribute otherwise. After resuming the first term of (\ref{prrrd}), we see, that $t^n \phi^{n+1}(t)$ generates the following distribution
\begin{eqnarray}
	\rho_{n}(x)&=&\rho_0(x)2^n x^n\frac{1}{2\sqrt{1-\frac{1}{x^2}}}\slr{\lr{1+\sqrt{1-\frac{1}{x^2}}}^{n+1}-\lr{1-\sqrt{1-\frac{1}{x^2}}}^{n+1}}=\nonumber\\&=&
	\frac{2^n}{i\pi}\slr{\lr{x+i\sqrt{1-x^2}}^{n+1}-\lr{x-i\sqrt{1-x^2}}^{n+1}}.
\end{eqnarray}
Since $|x|\leq1$, we can write $x=\cos\theta$ for some $\theta\in[0,\pi]$ and $x\pm i\sqrt{1-x^2}=e^{\pm i \theta}$. This yields
\begin{eqnarray}
	\rho_n(x)=\frac{2^n e^{i(n+1)\theta}-2^n e^{-i(n+1)\theta}}{i\pi}.
\end{eqnarray}
And the final distribution is then
\begin{eqnarray}
	\rho(x)=\sum_{n=0}^\infty a_n \rho_n(x)=\frac{1}{i\pi}\lr{e^{i\theta}f(2e^{i\theta})-e^{-i\theta}f(2e^{-i\theta})},
\end{eqnarray}
which is the desired formula (\ref{panati2}) for the one dimensional case.

\section{Explicit formulas for three and four matrix distributions}\label{explicit}

The factor $\tilde\rho_3(x,y,z)$ multiplying $\rho(x)\rho(y)\rho(z)$ in formula (\ref{rho3}) is given by fraction with the following numerator
\begin{eqnarray*}
	1- g_{12}^2- g_{13}^2+ g_{12}^2 g_{13}^2- g_{12} g_{13} g_{23}+ g_{12}^3 g_{13} g_{23}+ g_{12} g_{13}^3 g_{23}- g_{12}^3 g_{13}^3 g_{23}	- g_{23}^2+ g_{12}^2 g_{23}^2+\\
	+ g_{13}^2 g_{23}^2- g_{12}^2 g_{13}^2 g_{23}^2+ g_{12} g_{13} g_{23}^3- g_{12}^3 g_{13} g_{23}^3-    g_{12} g_{13}^3 g_{23}^3 +  g_{12}^3  g_{13} ^3   g_{23} ^3 - 
				4   g_{12} ^2   g_{13} ^2  x ^2+\\
	+  4 g_{12} g_{13} g_{23} x^2 + 4  g_{12} ^2   g_{13} ^2   g_{23} ^2  x ^2 -   4 g_{12} g_{13} g_{23}^3 x^2 + 4   g_{12}   g_{13} ^2  x  y - 4   g_{13}   g_{23}  x  y - 
  			4 g_{12}^2 g_{13} g_{23} x y+ \\
  + 4   g_{12} ^2   g_{13} ^3   g_{23}  x  y + 4   g_{12}   g_{23} ^2  x  y -  4 g_{12} g_{13}^2 g_{23}^2 x  y - 4   g_{12} ^3   g_{13} ^2   g_{23} ^2  x  y + 
  			  4 g_{12}^2 g_{13} g_{23}^3 x  y + 4   g_{12}   g_{13}   g_{23}  y ^2 - \\
  4 g_{12} g_{13}^3 g_{23} y^2 - 4   g_{12} ^2   g_{23} ^2  y ^2 +   4 g_{12}^2 g_{13}^2  g_{23}^2  y ^2 + 4   g_{12} ^2   g_{13}  x  z - 4   g_{12}   g_{23}  x  z - 4 g_{12} g_{13}^2 g_{23} x z\\
   + 4   g_{12} ^3   g_{13} ^2   g_{23}  x  z + 4   g_{13}   g_{23} ^2  x  z - 
  4 g_{12}^2 g_{13} g_{23}^2 x  z - 4   g_{12} ^2   g_{13} ^3   g_{23} ^2  x  z + 4 g_{12} g_{13}^2 g_{23}^3 x  z - 4   g_{12}   g_{13}  y  z \\
  + 4   g_{12} ^2   g_{23}  y  z + 
  4 g_{13}^2 g_{23}y z - 4   g_{12} ^2   g_{13} ^2   g_{23}  y  z - 4   g_{12}   g_{13}   g_{23} ^2  y  z + 4 g_{12}^3 g_{13} g_{23}^2 y  z \\
  + 4   g_{12}   g_{13} ^3   g_{23} ^2  y  z - 
  4 g_{12}^2 g_{13}^2  g_{23}^3  y  z + 4   g_{12}   g_{13}   g_{23}  z ^2 - 
  4 g_{12}^3 g_{13} g_{23} z^2 - 4   g_{13} ^2   g_{23} ^2  z ^2 + 
  4 g_{12}^2 g_{13}^2  g_{23}^2  z ^2
\end{eqnarray*}
and denominator
\[1-2\gamma_{12}^2+\gamma_{12}^4-2\gamma_{13}^2+4\gamma_{12}^2\gamma_{13}^2-2\gamma_{12}^4\gamma_{13}^2+\gamma_{13}^4-2\gamma_{12}^2\gamma_{13}^4+\gamma_{12}^4\gamma_{13}^4-2\gamma_{23}^2+4\gamma_{12}^2\gamma_{23}^2-2\gamma_{12}^4\gamma_{23}^2+4\gamma_{13}^2\gamma_{23}^2\]
\[-8\gamma_{12}^2\gamma_{13}^2\gamma_{23}^2+4\gamma_{12}^4\gamma_{13}^2\gamma_{23}^2-2\gamma_{13}^4\gamma_{23}^2+4\gamma_{12}^2\gamma_{13}^4\gamma_{23}^2-2\gamma_{12}^4\gamma_{13}^4\gamma_{23}^2+\gamma_{23}^4-2\gamma_{12}^2\gamma_{23}^4+\gamma_{12}^4\gamma_{23}^4-2\gamma_{13}^2\gamma_{23}^4\]
\[+4\gamma_{12}^2\gamma_{13}^2\gamma_{23}^4
-2\gamma_{12}^4\gamma_{13}^2\gamma_{23}^4+\gamma_{13}^4\gamma_{23}^4-2\gamma_{12}^2\gamma_{13}^4\gamma_{23}^4+\gamma_{12}^4\gamma_{13}^4\gamma_{23}^4+4\gamma_{12}^2x^2+4\gamma_{13}^2x^2-16\gamma_{12}^2\gamma_{13}^2x^2+4\gamma_{12}^4\gamma_{13}^2x^2\]
\[+4\gamma_{12}^2\gamma_{13}^4x^2-8\gamma_{12}^2\gamma_{23}^2x^2
-8\gamma_{13}^2\gamma_{23}^2x^2+32\gamma_{12}^2\gamma_{13}^2\gamma_{23}^2x^2-8\gamma_{12}^4\gamma_{13}^2\gamma_{23}^2x^2-8\gamma_{12}^2\gamma_{13}^4\gamma_{23}^2x^2+4\gamma_{12}^2\gamma_{23}^4x^2\]
\[+4\gamma_{13}^2\gamma_{23}^4x^2-16\gamma_{12}^2\gamma_{13}^2\gamma_{23}^4x^2+4\gamma_{12}^4\gamma_{13}^2\gamma_{23}^4x^2+4\gamma_{12}^2\gamma_{13}^4\gamma_{23}^4x^2+16\gamma_{12}^2\gamma_{13}^2x^4-32\gamma_{12}^2\gamma_{13}^2\gamma_{23}^2x^4+16\gamma_{12}^2\gamma_{13}^2\gamma_{23}^4x^4\]
\[-4\gamma_{12}xy-4\gamma_{12}^3xy+8\gamma_{12}\gamma_{13}^2xy+8\gamma_{12}^3\gamma_{13}^2xy-4\gamma_{12}\gamma_{13}^4xy
-4\gamma_{12}^3\gamma_{13}^4xy+8\gamma_{12}\gamma_{23}^2xy+8\gamma_{12}^3\gamma_{23}^2xy\]
\[-16\gamma_{12}\gamma_{13}^2\gamma_{23}^2xy-16\gamma_{12}^3\gamma_{13}^2\gamma_{23}^2xy+8\gamma_{12}\gamma_{13}^4\gamma_{23}^2xy+8\gamma_{12}^3\gamma_{13}^4\gamma_{23}^2xy-4\gamma_{12}\gamma_{23}^4xy-
4\gamma_{12}^3\gamma_{23}^4xy+8\gamma_{12}\gamma_{13}^2\gamma_{23}^4xy\]
\[+8\gamma_{12}^3\gamma_{13}^2\gamma_{23}^4xy-4\gamma_{12}\gamma_{13}^4\gamma_{23}^4xy-4\gamma_{12}^3\gamma_{13}^4\gamma_{23}^4xy-16\gamma_{12}\gamma_{13}^2x^3y-16\gamma_{12}^3\gamma_{13}^2x^3y+32\gamma_{12}\gamma_{13}^2\gamma_{23}^2x^3y\]
\[+32\gamma_{12}^3\gamma_{13}^2\gamma_{23}^2x^3y-16\gamma_{12}\gamma_{13}^2\gamma_{23}^4x^3y-16\gamma_{12}^3\gamma_{13}^2\gamma_{23}^4x^3y+4\gamma_{12}^2y^2-8\gamma_{12}^2\gamma_{13}^2y^2+4\gamma_{12}^2\gamma_{13}^4y^2+4\gamma_{23}^2y^2-16\gamma_{12}^2\gamma_{23}^2y^2\]
\[+4\gamma_{12}^4\gamma_{23}^2y^2-8\gamma_{13}^2\gamma_{23}^2y^2+32\gamma_{12}^2\gamma_{13}^2\gamma_{23}^2y^2-8\gamma_{12}^4\gamma_{13}^2\gamma_{23}^2y^2+4\gamma_{13}^4\gamma_{23}^2y^2-16\gamma_{12}^2\gamma_{13}^4\gamma_{23}^2y^2+4\gamma_{12}^4\gamma_{13}^4\gamma_{23}^2y^2\]
\[+4\gamma_{12}^2\gamma_{23}^4y^2
-8\gamma_{12}^2\gamma_{13}^2\gamma_{23}^4y^2+4\gamma_{12}^2\gamma_{13}^4\gamma_{23}^4y^2+16\gamma_{12}^2\gamma_{13}^2x^2y^2+16\gamma_{12}^2\gamma_{23}^2x^2y^2+16\gamma_{13}^2\gamma_{23}^2x^2y^2-96\gamma_{12}^2\gamma_{13}^2\gamma_{23}^2x^2y^2+\]
\[16\gamma_{12}^4\gamma_{13}^2\gamma_{23}^2x^2y^2+
16\gamma_{12}^2\gamma_{13}^4\gamma_{23}^2x^2y^2+
16\gamma_{12}^2\gamma_{13}^2\gamma_{23}^4x^2y^2+
64\gamma_{12}^2\gamma_{13}^2\gamma_{23}^2x^4y^2
-16\gamma_{12}\gamma_{23}^2xy^3
-16\gamma_{12}^3\gamma_{23}^2xy^3\]
\[+32\gamma_{12}\gamma_{13}^2\gamma_{23}^2xy^3
+32\gamma_{12}^3\gamma_{13}^2\gamma_{23}^2xy^3
-16\gamma_{12}\gamma_{13}^4\gamma_{23}^2xy^3
-16\gamma_{12}^3\gamma_{13}^4\gamma_{23}^2xy^3
-64\gamma_{12}\gamma_{13}^2\gamma_{23}^2x^3y^3
-64\gamma_{12}^3\gamma_{13}^2\gamma_{23}^2x^3y^3\]
\[+16\gamma_{12}^2\gamma_{23}^2y^4
-32\gamma_{12}^2\gamma_{13}^2\gamma_{23}^2y^4
+16\gamma_{12}^2\gamma_{13}^4\gamma_{23}^2y^4+
64\gamma_{12}^2\gamma_{13}^2\gamma_{23}^2x^2y^4
-4\gamma_{13}xz+8\gamma_{12}^2\gamma_{13}xz
-4\gamma_{12}^4\gamma_{13}xz-4\gamma_{13}^3xz\]
\[+8\gamma_{12}^2\gamma_{13}^3xz
-4\gamma_{12}^4\gamma_{13}^3xz
+8\gamma_{13}\gamma_{23}^2xz
-16\gamma_{12}^2\gamma_{13}\gamma_{23}^2xz
+8\gamma_{12}^4\gamma_{13}\gamma_{23}^2xz
+8\gamma_{13}^3\gamma_{23}^2xz-16\gamma_{12}^2\gamma_{13}^3\gamma_{23}^2xz\]
\[+8\gamma_{12}^4\gamma_{13}^3\gamma_{23}^2xz
-4\gamma_{13}\gamma_{23}^4xz
+8\gamma_{12}^2\gamma_{13}\gamma_{23}^4xz
-4\gamma_{12}^4\gamma_{13}\gamma_{23}^4xz
-4\gamma_{13}^3\gamma_{23}^4xz
+8\gamma_{12}^2\gamma_{13}^3\gamma_{23}^4xz\]
\[-4\gamma_{12}^4\gamma_{13}^3\gamma_{23}^4xz
-16\gamma_{12}^2\gamma_{13}x^3z
-16\gamma_{12}^2\gamma_{13}^3x^3z
+32\gamma_{12}^2\gamma_{13}\gamma_{23}^2x^3z
+32\gamma_{12}^2\gamma_{13}^3\gamma_{23}^2x^3z
-16\gamma_{12}^2\gamma_{13}\gamma_{23}^4x^3z\]
\[-16\gamma_{12}^2\gamma_{13}^3\gamma_{23}^4x^3z
-4\gamma_{23}yz+8\gamma_{12}^2\gamma_{23}yz
-4\gamma_{12}^4\gamma_{23}yz
+8\gamma_{13}^2\gamma_{23}yz
-16\gamma_{12}^2\gamma_{13}^2\gamma_{23}yz
+8\gamma_{12}^4\gamma_{13}^2\gamma_{23}yz-4\gamma_{13}^4\gamma_{23}yz\]
\[+8\gamma_{12}^2\gamma_{13}^4\gamma_{23}yz
-4\gamma_{12}^4\gamma_{13}^4\gamma_{23}yz
-4\gamma_{23}^3yz
+8\gamma_{12}^2\gamma_{23}^3yz
-4\gamma_{12}^4\gamma_{23}^3yz
+8\gamma_{13}^2\gamma_{23}^3yz
-16\gamma_{12}^2\gamma_{13}^2\gamma_{23}^3yz\]
\[+8\gamma_{12}^4\gamma_{13}^2\gamma_{23}^3yz
-4\gamma_{13}^4\gamma_{23}^3yz
+8\gamma_{12}^2\gamma_{13}^4\gamma_{23}^3yz
-4\gamma_{12}^4\gamma_{13}^4\gamma_{23}^3yz
+16\gamma_{12}\gamma_{13}x^2yz+
16\gamma_{12}^3\gamma_{13}x^2yz\]
\[+16\gamma_{12}\gamma_{13}^3x^2yz
+16\gamma_{12}^3\gamma_{13}^3x^2yz
-16\gamma_{12}^2\gamma_{23}x^2yz
-16\gamma_{13}^2\gamma_{23}x^2yz
+64\gamma_{12}^2\gamma_{13}^2\gamma_{23}x^2yz
-16\gamma_{12}^4\gamma_{13}^2\gamma_{23}x^2yz\]
\[-16\gamma_{12}^2\gamma_{13}^4\gamma_{23}x^2yz
-32\gamma_{12}\gamma_{13}\gamma_{23}^2x^2yz
-32\gamma_{12}^3\gamma_{13}\gamma_{23}^2x^2yz
-32\gamma_{12}\gamma_{13}^3\gamma_{23}^2x^2yz
-32\gamma_{12}^3\gamma_{13}^3\gamma_{23}^2x^2yz\]
\[-16\gamma_{12}^2\gamma_{23}^3x^2yz
-16\gamma_{13}^2\gamma_{23}^3x^2yz
+64\gamma_{12}^2\gamma_{13}^2\gamma_{23}^3x^2yz
-16\gamma_{12}^4\gamma_{13}^2\gamma_{23}^3x^2yz
-16\gamma_{12}^2\gamma_{13}^4\gamma_{23}^3x^2yz
+16\gamma_{12}\gamma_{13}\gamma_{23}^4x^2yz\]\[
+16\gamma_{12}^3\gamma_{13}\gamma_{23}^4x^2yz
+16\gamma_{12}\gamma_{13}^3\gamma_{23}^4x^2yz
+16\gamma_{12}^3\gamma_{13}^3\gamma_{23}^4x^2yz
-64\gamma_{12}^2\gamma_{13}^2\gamma_{23}x^4yz
-64\gamma_{12}^2\gamma_{13}^2\gamma_{23}^3x^4yz\]\[
-16\gamma_{12}^2\gamma_{13}xy^2z
-16\gamma_{12}^2\gamma_{13}^3xy^2z
+16\gamma_{12}\gamma_{23}xy^2z
+16\gamma_{12}^3\gamma_{23}xy^2z
-32\gamma_{12}\gamma_{13}^2\gamma_{23}xy^2z
-32\gamma_{12}^3\gamma_{13}^2\gamma_{23}xy^2z\]\[
+16\gamma_{12}\gamma_{13}^4\gamma_{23}xy^2z
+16\gamma_{12}^3\gamma_{13}^4\gamma_{23}xy^2z
-16\gamma_{13}\gamma_{23}^2xy^2z
+64\gamma_{12}^2\gamma_{13}\gamma_{23}^2xy^2z
-16\gamma_{12}^4\gamma_{13}\gamma_{23}^2xy^2z-
16\gamma_{13}^3\gamma_{23}^2xy^2z\]\[
+64\gamma_{12}^2\gamma_{13}^3\gamma_{23}^2xy^2z
-16\gamma_{12}^4\gamma_{13}^3\gamma_{23}^2xy^2z
+16\gamma_{12}\gamma_{23}^3xy^2z
+16\gamma_{12}^3\gamma_{23}^3xy^2z
-32\gamma_{12}\gamma_{13}^2\gamma_{23}^3xy^2z
-32\gamma_{12}^3\gamma_{13}^2\gamma_{23}^3xy^2z\]\[
+16\gamma_{12}\gamma_{13}^4\gamma_{23}^3xy^2z+
16\gamma_{12}^3\gamma_{13}^4\gamma_{23}^3xy^2z
-16\gamma_{12}^2\gamma_{13}\gamma_{23}^4xy^2z
-16\gamma_{12}^2\gamma_{13}^3\gamma_{23}^4xy^2z
+64\gamma_{12}\gamma_{13}^2\gamma_{23}x^3y^2z\]\[
+64\gamma_{12}^3\gamma_{13}^2\gamma_{23}x^3y^2z
-64\gamma_{12}^2\gamma_{13}\gamma_{23}^2x^3y^2z-
64\gamma_{12}^2\gamma_{13}^3\gamma_{23}^2x^3y^2z
+64\gamma_{12}\gamma_{13}^2\gamma_{23}^3x^3y^2z
+64\gamma_{12}^3\gamma_{13}^2\gamma_{23}^3x^3y^2z\]\[
-16\gamma_{12}^2\gamma_{23}y^3z
+32\gamma_{12}^2\gamma_{13}^2\gamma_{23}y^3z
-16\gamma_{12}^2\gamma_{13}^4\gamma_{23}y^3z
-16\gamma_{12}^2\gamma_{23}^3y^3z
+32\gamma_{12}^2\gamma_{13}^2\gamma_{23}^3y^3z
-16\gamma_{12}^2\gamma_{13}^4\gamma_{23}^3y^3z\]\[
-64\gamma_{12}^2\gamma_{13}^2\gamma_{23}x^2y^3z
+64\gamma_{12}\gamma_{13}\gamma_{23}^2x^2y^3z
+64\gamma_{12}^3\gamma_{13}\gamma_{23}^2x^2y^3z
+64\gamma_{12}\gamma_{13}^3\gamma_{23}^2x^2y^3z
+64\gamma_{12}^3\gamma_{13}^3\gamma_{23}^2x^2y^3z\]\[
-64\gamma_{12}^2\gamma_{13}^2\gamma_{23}^3x^2y^3z
-64\gamma_{12}^2\gamma_{13}\gamma_{23}^2xy^4z
-64\gamma_{12}^2\gamma_{13}^3\gamma_{23}^2xy^4z
+4\gamma_{13}^2z^2-8\gamma_{12}^2\gamma_{13}^2z^2
+4\gamma_{12}^4\gamma_{13}^2z^2\]\[
+4\gamma_{23}^2z^2-8\gamma_{12}^2\gamma_{23}^2z^2
+4\gamma_{12}^4\gamma_{23}^2z^2
-16\gamma_{13}^2\gamma_{23}^2z^2
+32\gamma_{12}^2\gamma_{13}^2\gamma_{23}^2z^2
-16\gamma_{12}^4\gamma_{13}^2\gamma_{23}^2z^2
+4\gamma_{13}^4\gamma_{23}^2z^2\]\[
-8\gamma_{12}^2\gamma_{13}^4\gamma_{23}^2z^2
+4\gamma_{12}^4\gamma_{13}^4\gamma_{23}^2z^2
+4\gamma_{13}^2\gamma_{23}^4z^2
-8\gamma_{12}^2\gamma_{13}^2\gamma_{23}^4z^2
+4\gamma_{12}^4\gamma_{13}^2\gamma_{23}^4z^2
+16\gamma_{12}^2\gamma_{13}^2x^2z^2\]\[
+16\gamma_{12}^2\gamma_{23}^2x^2z^2
+16\gamma_{13}^2\gamma_{23}^2x^2z^2
-96\gamma_{12}^2\gamma_{13}^2\gamma_{23}^2x^2z^2
+16\gamma_{12}^4\gamma_{13}^2\gamma_{23}^2x^2z^2
+16\gamma_{12}^2\gamma_{13}^4\gamma_{23}^2x^2z^2
+16\gamma_{12}^2\gamma_{13}^2\gamma_{23}^4x^2z^2\]\[
+64\gamma_{12}^2\gamma_{13}^2\gamma_{23}^2x^4z^2
-16\gamma_{12}\gamma_{13}^2xyz^2
-16\gamma_{12}^3\gamma_{13}^2xyz^2
+16\gamma_{13}\gamma_{23}xyz^2
-32\gamma_{12}^2\gamma_{13}\gamma_{23}xyz^2
+16\gamma_{12}^4\gamma_{13}\gamma_{23}xyz^2\]\[
+16\gamma_{13}^3\gamma_{23}xyz^2
-32\gamma_{12}^2\gamma_{13}^3\gamma_{23}xyz^2
+16\gamma_{12}^4\gamma_{13}^3\gamma_{23}xyz^2-
16\gamma_{12}\gamma_{23}^2xyz^2
-16\gamma_{12}^3\gamma_{23}^2xyz^2
+64\gamma_{12}\gamma_{13}^2\gamma_{23}^2xyz^2\]\[
+64\gamma_{12}^3\gamma_{13}^2\gamma_{23}^2xyz^2
-16\gamma_{12}\gamma_{13}^4\gamma_{23}^2xyz^2
-16\gamma_{12}^3\gamma_{13}^4\gamma_{23}^2xyz^2
+16\gamma_{13}\gamma_{23}^3xyz^2
-32\gamma_{12}^2\gamma_{13}\gamma_{23}^3xyz^2
+16\gamma_{12}^4\gamma_{13}\gamma_{23}^3xyz^2\]\[
+16\gamma_{13}^3\gamma_{23}^3xyz^2
-32\gamma_{12}^2\gamma_{13}^3\gamma_{23}^3xyz^2
+16\gamma_{12}^4\gamma_{13}^3\gamma_{23}^3xyz^2
-16\gamma_{12}\gamma_{13}^2\gamma_{23}^4xyz^2
-16\gamma_{12}^3\gamma_{13}^2\gamma_{23}^4xyz^2\]\[
+64\gamma_{12}^2\gamma_{13}\gamma_{23}x^3yz^2+
64\gamma_{12}^2\gamma_{13}^3\gamma_{23}x^3yz^2-
64\gamma_{12}\gamma_{13}^2\gamma_{23}^2x^3yz^2-
64\gamma_{12}^3\gamma_{13}^2\gamma_{23}^2x^3yz^2+
64\gamma_{12}^2\gamma_{13}\gamma_{23}^3x^3yz^2+\]\[
64\gamma_{12}^2\gamma_{13}^3\gamma_{23}^3x^3yz^2
+16\gamma_{12}^2\gamma_{13}^2y^2z^2+
16\gamma_{12}^2\gamma_{23}^2y^2z^2
+16\gamma_{13}^2\gamma_{23}^2y^2z^2-
96\gamma_{12}^2\gamma_{13}^2\gamma_{23}^2y^2z^2+\]\[
16\gamma_{12}^4\gamma_{13}^2\gamma_{23}^2y^2z^2+
16\gamma_{12}^2\gamma_{13}^4\gamma_{23}^2y^2z^2+
16\gamma_{12}^2\gamma_{13}^2\gamma_{23}^4y^2z^2
-64\gamma_{12}\gamma_{13}\gamma_{23}x^2y^2z^2-
64\gamma_{12}^3\gamma_{13}\gamma_{23}x^2y^2z^2-\]\[
64\gamma_{12}\gamma_{13}^3\gamma_{23}x^2y^2z^2-
64\gamma_{12}^3\gamma_{13}^3\gamma_{23}x^2y^2z^2+
128\gamma_{12}^2\gamma_{13}^2\gamma_{23}^2x^2y^2z^2-
64\gamma_{12}\gamma_{13}\gamma_{23}^3x^2y^2z^2-
64\gamma_{12}^3\gamma_{13}\gamma_{23}^3x^2y^2z^2-\]\[
64\gamma_{12}\gamma_{13}^3\gamma_{23}^3x^2y^2z^2-
64\gamma_{12}^3\gamma_{13}^3\gamma_{23}^3x^2y^2z^2+
64\gamma_{12}^2\gamma_{13}\gamma_{23}xy^3z^2+
64\gamma_{12}^2\gamma_{13}^3\gamma_{23}xy^3z^2
-64\gamma_{12}\gamma_{13}^2\gamma_{23}^2xy^3z^2-\]\[
64\gamma_{12}^3\gamma_{13}^2\gamma_{23}^2xy^3z^2+
64\gamma_{12}^2\gamma_{13}\gamma_{23}^3xy^3z^2+
64\gamma_{12}^2\gamma_{13}^3\gamma_{23}^3xy^3z^2+
64\gamma_{12}^2\gamma_{13}^2\gamma_{23}^2y^4z^2
-16\gamma_{13}\gamma_{23}^2xz^3\]\[
+32\gamma_{12}^2\gamma_{13}\gamma_{23}^2xz^3
-16\gamma_{12}^4\gamma_{13}\gamma_{23}^2xz^3-
16\gamma_{13}^3\gamma_{23}^2xz^3
+32\gamma_{12}^2\gamma_{13}^3\gamma_{23}^2xz^3-
16\gamma_{12}^4\gamma_{13}^3\gamma_{23}^2xz^3
-64\gamma_{12}^2\gamma_{13}\gamma_{23}^2x^3z^3-\]\[
64\gamma_{12}^2\gamma_{13}^3\gamma_{23}^2x^3z^3
-16\gamma_{13}^2\gamma_{23}yz^3+
32\gamma_{12}^2\gamma_{13}^2\gamma_{23}yz^3
-16\gamma_{12}^4\gamma_{13}^2\gamma_{23}yz^3-
16\gamma_{13}^2\gamma_{23}^3yz^3
+32\gamma_{12}^2\gamma_{13}^2\gamma_{23}^3yz^3-\]\[
16\gamma_{12}^4\gamma_{13}^2\gamma_{23}^3yz^3
-64\gamma_{12}^2\gamma_{13}^2\gamma_{23}x^2yz^3+
64\gamma_{12}\gamma_{13}\gamma_{23}^2x^2yz^3
+64\gamma_{12}^3\gamma_{13}\gamma_{23}^2x^2yz^3+
64\gamma_{12}\gamma_{13}^3\gamma_{23}^2x^2yz^3+\]\[
64\gamma_{12}^3\gamma_{13}^3\gamma_{23}^2x^2yz^3-
64\gamma_{12}^2\gamma_{13}^2\gamma_{23}^3x^2yz^3+
64\gamma_{12}\gamma_{13}^2\gamma_{23}xy^2z^3+64\gamma_{12}^3\gamma_{13}^2\gamma_{23}xy^2z^3-
64\gamma_{12}^2\gamma_{13}\gamma_{23}^2xy^2z^3-\]\[
64\gamma_{12}^2\gamma_{13}^3\gamma_{23}^2xy^2z^3+
64\gamma_{12}\gamma_{13}^2\gamma_{23}^3xy^2z^3+
64\gamma_{12}^3\gamma_{13}^2\gamma_{23}^3xy^2z^3-
64\gamma_{12}^2\gamma_{13}^2\gamma_{23}y^3z^3-64\gamma_{12}^2\gamma_{13}^2\gamma_{23}^3y^3z^3+
16\gamma_{13}^2\gamma_{23}^2z^4\]\[
-32\gamma_{12}^2\gamma_{13}^2\gamma_{23}^2z^4+
16\gamma_{12}^4\gamma_{13}^2\gamma_{23}^2z^4
+64\gamma_{12}^2\gamma_{13}^2\gamma_{23}^2x^2z^4-
64\gamma_{12}\gamma_{13}^2\gamma_{23}^2xyz^4-64\gamma_{12}^3\gamma_{13}^2\gamma_{23}^2xyz^4+
64\gamma_{12}^2\gamma_{13}^2\gamma_{23}^2y^2z^4\]

The factor $\tilde\rho_4(x,y,z,w)$ multiplying $\rho(x)\rho(y)\rho(z)\rho(w)$ in formula (\ref{rho4}) is given by
\[\frac{(1-\gamma^2)^3\slr{1+\gamma^6-\gamma^2(1+\gamma^2)(1+4xz+4yw)+4\gamma^3(x+z)(w+y)}}{denominator}\]
with the denominator
\[(1+\gamma^{2})^8-4\gamma(1+\gamma^{14})(x+z)(w+y)8\gamma^2(1+\gamma^{12})\Big[x^2+y^2+z^2+w^2+2(xy+zw)(xw+yz)+4xyzw\Big]\]\[
	-4\gamma^3(1+\gamma^{10})(x+z)(w+y)\Big[-5+4(x^2+y^2+z^2+w^2)+4(xz+wy)+16xyzw\Big]\]\[
	+16\gamma^4(1+\gamma^{8})\Big[
	-3(x^2+y^2+z^2+w^2)
	+(x^4+y^4+z^4+w^4)+3(x^2+z^2)(y^2+w^2)+4(x^2z^2+y^2w^2)\]\[-4xyzw
	-2(xy+zw)(xw+yz)+8xyzw(x^2+y^2+z^2+w^2)+4(x^2+z^2)(y^2+w^2)(xz+wy)\]\[+16xyzw(xz+wy)+16x^2y^2z^2w^2
	\Big]\]\[
	-4\gamma^5(1+\gamma^{6})(x+z)(w+y)\Big[9
	-12(x^2+y^2+z^2+w^2+xz+wy)+16(x^2+z^2)(y^2+w^2)\]\[
	+16(x^2z^2+y^2w^2)+16xz(x^2+z^2)+16yw(y^2+w^2)+48xyzw+64xyzw(xz+wy)\Big]\]\[
	8\gamma^6(1+\gamma^{4})\Big[15(x^2+y^2+z^2+w^2)-8(x^4+y^4+z^4+w^4)-2(xy+zw)(xw+yz)\]\[
	-24(x^2+z^2)(y^2+w^2)-32(x^2z^2+y^2w^2)-4xyzw+32(x^2y^2z^2+\textrm{perm})+16x^2z^2(x^2+z^2)\]\[
	+16y^2w^2(y^2+w^2)+8(x^2+z^2)(y^4+w^4)+8(x^4+z^4)(y^2+w^2)\]\[	+32xyzw(xz(x^2+z^2)+yw(y^2+w^2))+32xyzw(x^2+z^2)(y^2+w^2)\]\[+64xyzw(x^2z^2+y^2w^2)+64xyzw(xy+zw)(xw+yz)+32(x^3y^2z^3+\textrm{cycl})+128x^2y^2z^2w^2
	\Big]\]\[
	-4\gamma^7(1+\gamma^{2})(x+z)(w+y)\Big[-5+8(x^2+y^2+z^2+w^2)+8(xz+yw)\]\[
	-16(x^2+z^2)(y^2+w^2)-16xz(x^2+z^2+xz)-16yw(y^2+w^2+yw)-64xyzw\]\[
	+64xyzw(x^2+y^2+z^2+w^2)+128xyzw(xz+yw)+64(x^3z^3+y^3w^3)+64(x^2y^2z^2+\textrm{perm})
	\Big]\]\[
	32\gamma^8\Big[-5(x^2+y^2+z^2+w^2)+3(x^4+y^4+z^4+w^4)+9(x^2+z^2)(y^2+w^2)\]\[
	+2(xy+zw)(xw+yz)+12(x^2z^2+y^2w^2)+4xyzw-8x^2z^2(x^2+z^2)-8y^2w^2(y^2+z^2)\]\[
	-4(x^2+z^2)(y^4+w^4)-4(x^4+z^4)(y^2+w^2)-8xyzw(x^2+y^2+z^2+w^2)-16xyzw(xz+yw)\]\[
	-16(x^2y^2z^2+\textrm{perm})-4(x^2+z^2)(y^2+w^2)(xz+wy)+8(x^4z^4+y^4w^4)+64x^2y^2z^2w^2\]\[	+32xyzw(x^2z^2+y^2w^2)+16xyzw(x^2+z^2)(y^2+w^2)+16xyzw(xz(x^2+z^2)+yw(y^2+w^2))\]\[+32xyzw(xy+zw)(xw+yz)\\+16(x^3y^2z^3+\textrm{cycl})+8(x^2y^2z^4+x^2y^4z^2+x^4y^2z^2+\textrm{perm}))
	\Big]\]

Note that all the polynomials have the desired symmetry $x\leftrightarrow z, y\leftrightarrow w,(x,y)\leftrightarrow(z,w)$.

\end{document}